\documentclass[twocolumn,showpacs,preprintnumbers,amsmath,amssymb]{revtex4}
\usepackage{color}
\usepackage{graphicx}
\usepackage{dcolumn}
\usepackage{bm}

\begin{document}
\title{Antiferromagnetic transitions in EuFe$_2$As$_2$: A possible parent compound for superconductors}
\author{Zhi Ren,$^{1}$ Zengwei Zhu,$^{1}$ Shuai Jiang,$^{1}$ Xiangfan Xu,$^{1}$ Qian Tao,$^{1}$ Cao Wang,$^{1}$ Chunmu Feng,$^{2}$ Guanghan Cao$^{1,}$\footnote[1]{Electronic address: ghcao@zju.edu.cn} and Zhu'an Xu$^{1,}$\footnote[2]{Electronic address: zhuan@zju.edu.cn}}

\affiliation{$^{1}$Department of Physics, Zhejiang University, Hangzhou 310027, People's Republic of China}
\affiliation{$^{2}$Test and Analysis Center, Zhejiang University, Hangzhou 310027, People's Republic of China}
\date{\today}

\begin{abstract}
Ternary iron arsenide EuFe$_2$As$_2$ with ThCr$_2$Si$_2$-type structure has been studied by magnetic
susceptibility, resistivity, thermopower, Hall and specific heat measurements. The compound undergoes two
magnetic phase transitions at about 200 K and 20 K, respectively. The former was found to be accompanied with a
slight drop in magnetic susceptibility (after subtracting the Curie-Weiss paramagnetic contribution), a rapid
decrease in resistivity, a large jump in thermopower and a sharp peak in specific heat with decreasing
temperature, all of which point to a spin-density-wave-like antiferromagnetic transition. The latter was
proposed to be associated with an A-type antiferromagnetic ordering of Eu$^{2+}$ moments. Comparing with the
physical properties of the iso-structural compounds BaFe$_2$As$_2$ and SrFe$_2$As$_2$, we expect that
superconductivity could be induced in EuFe$_2$As$_2$ through appropriate doping.
\end{abstract}


\maketitle

Superconductivity in iron-based oxyarsenides has recently become a hot topic in condensed matter physics
community. The excitement was initiated by the observation of 26-K superconductivity in
LaFeAsO$_{1-x}$F$_{x}$\cite{Hosono}. Following this discovery, $T_c$ was raised quickly above 40 K by replacing
La with other lanthanides\cite{Chen-Sm,WNL-Ce,Ren-Pr,Ren-Nd}. Recently, $T_c$ has achieved 56 K in Th$^{4+}$
doped GdFeAsO.\cite{Wang-Th} These $T_c$ values surpass any other superconductors except high-$T_c$ cuprates,
indicating the emergence of a new class of high temperature superconductors. The parent compounds of these
superconductors LnFeAsO (Ln=lanthanides) adopt the ZrCuSiAs-type structure\cite{Johnson&Jeitschko,Quebe}, which
consists of alternating stacking [Ln$_{2}$O$_{2}$]$^{2+}$ and [Fe$_{2}$As$_{2}$]$^{2-}$ layers. X-ray and
neutron diffraction studies indicate that the parent compound LaFeAsO undergoes a structural phase transition
from tetragonal to orthorhombic at 155 K\cite{DaiPC Neutron,Hosono syncho}, followed by a spin-density-wave
(SDW) antiferromagnetic (AFM) transition below 137 K.\cite{DaiPC Neutron} The electron doping onto the
[Fe$_{2}$As$_{2}$]$^{2-}$ layers via heterovalent chemical substitution at [Ln$_{2}$O$_{2}$]$^{2+}$ layers
suppresses the phase transitions and induces superconductivity. Hence the iron-based oxyarsenide superconductors
provide a new platform to study the interplay between magnetism and superconductivity, as shown by many
theoretical studies.\cite{Kotliar,Cao,Singh}

Ternary iron arsenides $A$Fe$_2$As$_2$ ($A$=Sr,Ba)\cite{Pfisterer1980,Pfisterer1983} crystallize in
ThCr$_{2}$Si$_{2}$-type structure, which is built up with identical [Fe$_2$As$_2$]$^{2-}$ layers separated by
$A^{2+}$ instead of [Ln$_{2}$O$_{2}$]$^{2+}$ layers. In analogy with LnFeAsO, $A$Fe$_2$As$_2$ undergoes a
structural phase transition with a symmetry reduction from tetragonal to orthorhombic, accompanied by the
anomalies in electrical resistivity, magnetic susceptibility and specific heat.\cite{BaFe2As2,SrFe2As2}
Substitution of K$^{+}$ for $A^{2+}$ suppresses the phase transition and results in the occurrence of
superconductivity at about 38 K.\cite{BaFe2As2 SC,SrFe2As2 SC,ChenXH}

EuFe$_2$As$_2$ is another member of the ternary iron arsenide family,\cite{EuFeAs} however, only few work was
performed on this material. M\"{o}ssbauer and magnetic susceptibility studies\cite{Mossbuar} indicated that
EuFe$_2$As$_2$ experienced two magnetic transitions. The first one around 200 K was due to the AFM transition in
the iron sublattice. The second one at 19 K arose from the AFM ordering of Eu$^{2+}$ magnetic moments. No other
physical properties of EuFe$_2$As$_2$ have been reported. In order to assess the potential of inducing
superconductivity in this compound, we have carried out a systematic study of the physical properties of
EuFe$_2$As$_2$. We found that the transition at about 200 K was accompanied by a rapid decrease in resistivity,
a large jump in thermopower and a sharp peak in specific heat. In addition, a slight drop in magnetic
susceptibility was observed after subtracting the Curie-Weiss paramagnetic contribution of Eu$^{2+}$ magnetic
moments. These properties are quite similar with those of BaFe$_2$As$_2$ and SrFe$_2$As$_2$, suggesting that
EuFe$_2$As$_2$ is another possible parent compound in which superconductivity may be found by proper doping.

Polycrystalline samples of EuFe$_2$As$_2$ were synthesized from stoichiometric amounts of the elements as
reported previously\cite{EuFeAs}. Fresh Eu grains, Fe powders and As grains were mixed in a ratio of 1:2:2,
sealed in an evacuated quartz tube and sintered at 773 K for 12 hours then 1073 K for another 12 hours. After
cooling, the reaction product was thoroughly ground in an agate mortar and pressed into pellets under a pressure
of 2000 kg/cm$^{2}$ in an argon-filled glove-box. The pellets were annealed in an evacuated quartz tube at 1123
K for 12 hours and furnace-cooled to room temperature. The EuFe$_2$As$_2$ samples were obtained as black
powders, which is stable in air.

Powder X-ray diffraction (XRD) was performed at room temperature using a D/Max-rA diffractometer with
Cu-K$_{\alpha}$ radiation and a graphite monochromator. Lattice parameters were refined by a least-squares fit
using at least 15 XRD peaks. The electrical resistivity was measured using a standard four-probe method. The
temperature dependence of dc magnetization was measured on a Quantum Design Magnetic Property Measurement System
(MPMS-5). Resistivity, Hall, specific heat measurements were performed on a Quantum Design Physical Property
Measurement System (PPMS-9). Thermopower measurements were carried out in a cryogenic refrigerator down to 14 K
by a steady-state technique with a temperature gradient $\sim$ 1 K/cm.

Figure 1 shows an XRD pattern for the EuFe$_2$As$_2$ sample. Most of the diffraction peaks can be indexed based
on the ThCr$_{2}$Si$_{2}$-type structure. The refined lattice parameters are $a$=3.9104 {\AA} and $c$=12.1362
{\AA}, in agreement with the previous report\cite{EuFeAs}. Small amount of FeAs impurity was also observed in
the XRD pattern, which may arise from the loss of Eu during the high temperature sintering process.

\begin{figure}
\includegraphics[width=8.5cm]{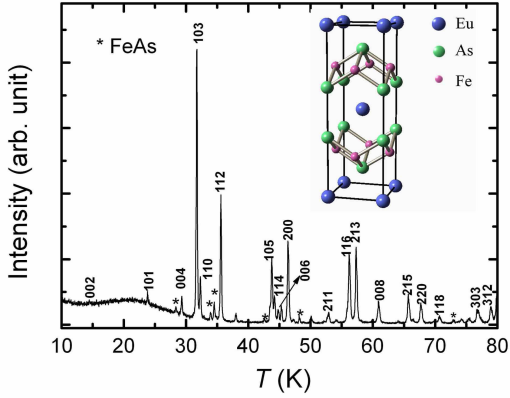}
\caption{X-ray powder diffraction pattern at room temperature for the EuFe$_2$As$_2$ sample. Small amount of
FeAs impurity was identified. The inset shows schematic crystal structure of EuFe$_2$As$_2$.}
\end{figure}

\begin{figure}
\includegraphics[width=8cm]{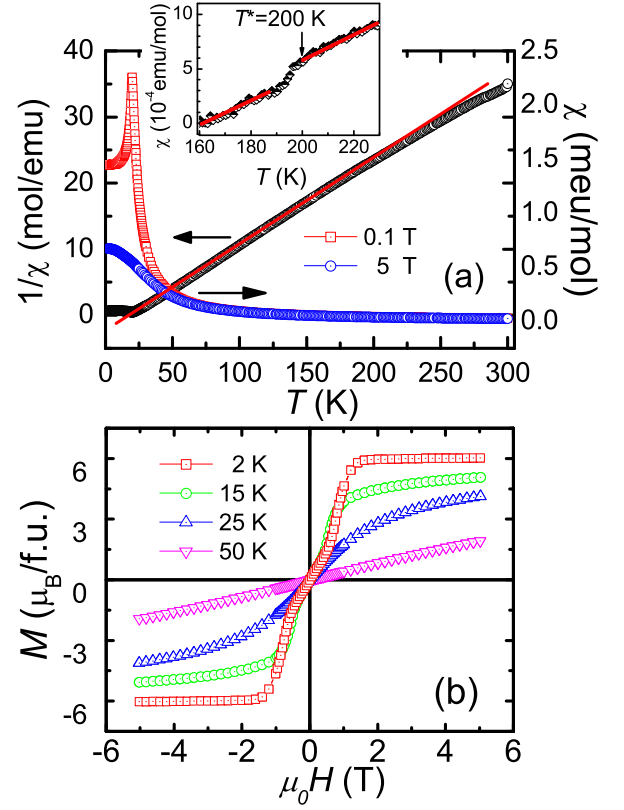}
\caption{(color online) (a) Temperature dependence of magnetic susceptibility for the EuFe$_2$As$_2$ sample. The
inset shows a drop in $\chi$ at $T$*=200 K, after subtraction of the Curie-Weiss contribution of Eu$^{2+}$
moments. The red lines are the guide to the eyes. (b) Field dependence of magnetization at low temperatures for
the EuFe$_2$As$_2$ sample.}
\end{figure}

Figure 2(a) shows magnetic susceptibility ($\chi$) measurement on the EuFe$_2$As$_2$ sample. The data of 20 K
$<T<$ 200 K obeys the Curie-Weiss law,
\begin{equation}
\frac{1}{\chi}=\frac{T+\theta}{C},
\end{equation}
where $C$ denotes the Curie-Weiss constant and $\theta$ the Weiss temperature. The fitted parameters are
$C$=7.58 emu$\cdot$K/mol and $\theta$=$-$19.7 K. The calculated effective magnetic moment is 7.79 $\mu_{B}$ per
formula unit, which is close to the theoretical value of 7.94 $\mu_{B}$ for a free Eu$^{2+}$ ion. The $\chi$
data above 200 K deviates slightly from the Curie-Weiss formula, as can be seen in the plot of 1/$\chi$ versus
$T$. After subtraction of the above Curie-Weiss contribution, a small drop in $\chi$ at 200 K can be found. This
behavior resembles those observed in BaFe$_2$As$_2$\cite{BaFe2As2} and SrFe$_2$As$_2$\cite{SrFe2As2}, which were
ascribed to an SDW transition. Below 20 K, $\chi$ decreases sharply, consistent with the previous
report\cite{Mossbuar}. $^{151}$Eu M\"{o}ssbauer spectroscopy indicated that the transition was due to the AFM
ordering of Eu$^{2+}$ moments. However, the Neel transition disappears under strong magnetic field.

Figure 2(b) shows the field-dependent magnetization for the EuFe$_2$As$_2$ sample at various temperatures. A
slope change in the $M-H$ curve at 2 K can be seen clearly at $\mu_{0}H$ = 0.65 T. In the range of 0.65 T $<
\mu_{0}H<$ 1.0 T, the magnetization increases rapidly with the field, suggesting a metamagnetic transition. The
magnetization saturates for $\mu_{0}H\geq$ 1.6 T, corresponding to an effective magnetic moment of 6
$\mu_{B}$/f.u. Though this value was underestimated owing to the presence of small amount of FeAs impurity, the
actual saturated moment would be still smaller than the expected value of 7 (=g$S$) $\mu_{B}$/f.u. This
deviation of magnitude of Eu$^{2+}$ moments was observed in Eu metal\cite{Eu} and
EuZn$_{2}$Sb$_{2}$\cite{EuZn2Sb2}, which may be due to the crystal field interactions.  The metamagnetic
transition reminisces the A-type antiferromagnetism in layered systems, such as
La$_{2-x}$Sr$_{1+x}$Mn$_{2}$O$_{7}$\cite{Kimura} and Na$_{0.85}$CoO$_{2}$\cite{Luo}. Note that the above
Curie-Weiss fit gives negative value of $\theta$, which means ferromagnetic interaction among the Eu$^{2+}$
moments. We propose that the AFM ordering of Eu$^{2+}$ spins is of A-type, \emph{i. e.}, the Eu$^{2+}$ moments
parallel in $ab$-plane but antiparallel along $c$-axis. Further study is needed to understand this interesting
issue.

\begin{figure}
\includegraphics[width=8cm]{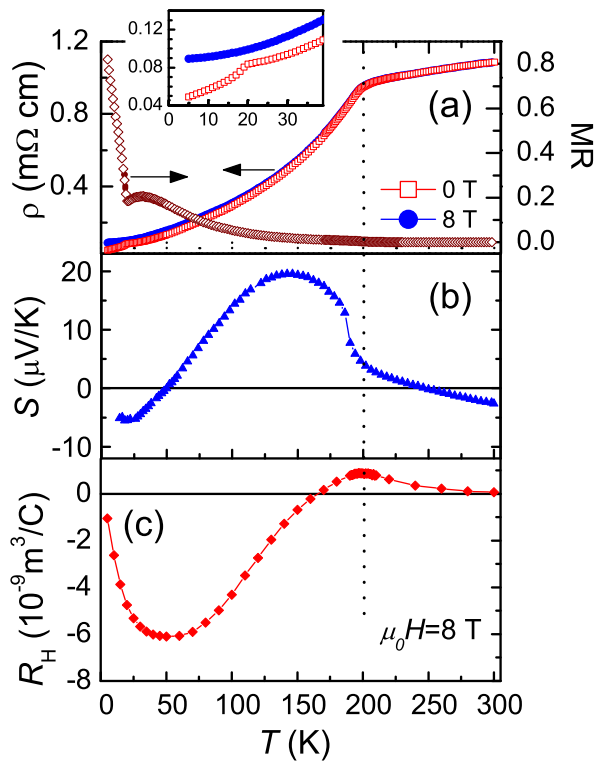}
\caption{(color online) Temperature dependence of (a) resistivity, (b) thermopower and (c) Hall coefficient for
the EuFe$_2$As$_2$ sample. The inset of (a) is an expanded plot for showing low-temperature data.}
\end{figure}

The transport properties of the EuFe$_2$As$_2$ sample are summarized in Figure 3. All these data demonstrate an
anomaly at 200 K, in correspondence with the above magnetic measurements. The resistivity ($\rho$) value at room
temperature is $\sim$ 1 m$\Omega$$\cdot$cm, comparable to those of SrFe$_2$As$_2$ \cite{SrFe2As2} and
BaFe$_2$As$_2$\cite{BaFe2As2} but much smaller than that of LaFeAsO\cite{WNL SDW}. With decreasing temperature,
$\rho$ decreases slightly. Below 200 K, a sudden slope change appears in the $\rho$-$T$ curve. Around 20 K, a
kink is evident in the inset of Fig. 3(a), which is related to the AFM ordering of Eu$^{2+}$ ions. No
resistivity anomaly can be detected around 77 K (the AFM transition temperature of FeAs\cite{FeAs}), indicating
no significant effect of the FeAs impurity on the resistivity behavior. Under a magnetic field of 8 T, there is
no observable variation in $\rho$ above 200 K. Below 200 K, however, we observed a positive magnetoresistance
(MR), which increases with decreasing temperature. This phenomenon can be explained in terms of the enhancement
of spin scattering due to the suppression of AFM SDW order by external magnetic field, similar to that in
LaFeAsO\cite{WNL SDW}. Below 19 K, the MR increases steeply and achieves 80\% at 5 K. We also note that the
resistivity kink at 19 K under zero field is totally suppressed by the strong field. This phenomenon coincides
with the magnetic structure proposed above because the interlayer AFM coupling is relatively weak and can be
easily destroyed by external magnetic field.

Figure 3(b) plots the thermopower ($S$) as a function of temperature. With decreasing temperature, $S$ is seen
to change its sign from being negative to being positive, then to being negative again. This behavior strongly
suggests multi-band characteristic in EuFe$_2$As$_2$. In Figure 3(c), a sign reversal in Hall coefficient
($R_{H}$) with decreasing temperature was also observed, further supporting the multi-band scenario. At 200 K, a
jump in $S$ and a turning point in $R_{H}$ are clearly seen, suggesting a substantial change of electronic
states at the Fermi level. The positive sign of $R_{H}$ above 200 K indicates that the dominant carriers are
hole-like, which is in contrast with that in LaFeAsO\cite{WNL SDW}. It is noted here that the Hall voltage is
nonlinear with the applied field below 50 K, which is probably related to the nonlinearity in the above $M-H$
curves. Thus the values of $R_{H}$ below 50 K are not well defined.

\begin{figure}
\includegraphics[width=8cm]{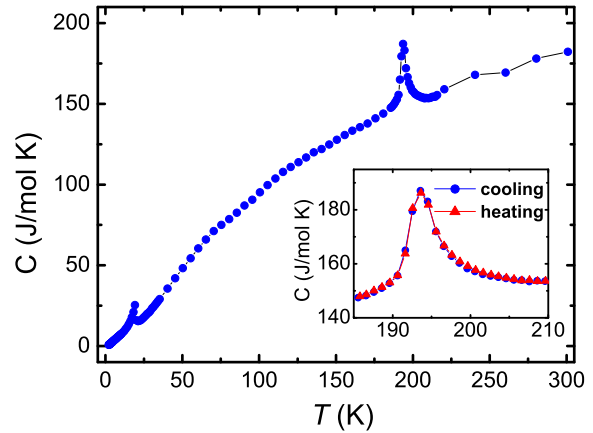}
\caption{(color online) Temperature dependence of specific heat for the EuFe$_2$As$_2$ sample. The inset is an
expanded plot, showing no detectable thermal hysteresis around 200 K.}
\end{figure}

Figure 4 shows the specific heat as a function of temperature. A sharp peak appears at 200 K, which points to a
first-order phase transition. However, no thermal hysteresis was found within the experimental precision. We
thus speculate that the latent heat is quite small. Another specific heat anomaly at 19 K is due to the AFM
ordering of Eu$^{2+}$ moments. Because of the influence of this transition, the electronic specific heat
coefficient $\gamma$ cannot be extracted reliably.

In summary, we have performed a systematic physical property measurements on EuFe$_2$As$_2$ polycrystalline
sample. It is found that the magnetic transition at 200 K is very similar to those of BaFe$_2$As$_2$ and
SrFe$_2$As$_2$, both of which become superconductors upon doping with some alkaline elements\cite{BaFe2As2
SC,SrFe2As2 SC,ChenXH}. Therefore, superconductivity may be induced by suppressing the SDW transition through
appropriate doping in EuFe$_2$As$_2$. The Eu$^{2+}$ moments in EuFe$_2$As$_2$ are expected to be compatible with
superconductivity, in analogy with CeFeAsO$_{1-x}$F$_{x}$\cite{WNL-Ce}. Hall measurements\cite{SrFe2As2
SC,ChenXH} indicated that the superconductivity in K-doped BaFe$_2$As$_2$ and SrFe$_2$As$_2$ was induced by hole
doping. Meanwhile, attempts to introduce superconductivity in BaFe$_2$As$_2$ by electron doping through
substitution of La$^{3+}$ for Ba$^{2+}$ were not successful.\cite{ChenXH} Owing to the dominant hole-like
carriers in EuFe$_2$As$_2$, we speculate that superconductivity may also be realized by hole-doping in
EuFe$_2$As$_2$ systems.

\begin{acknowledgments}
The authors thank Ying Liu for helpful discussions. This work is supported by the National Basic Research
Program of China (No.2006CB601003 and 2007CB925001) and the PCSIRT of the Ministry of Education of China
(IRT0754). \end{acknowledgments}

\end{document}